**Analog Circuit Applications based on Ambipolar Graphene/MoTe$_2$ Vertical Transistors**

*Chen Pan†, Yajun Fu†, Jiaxin Wang, Junwen Zeng, Guangxu Su, Mingsheng Long, Erfu Liu, Chenyu Wang, Anyuan Gao, Miao Wang, Yu Wang, Zhenlin Wang, Shi-Jun Liang\*, Ru Huang, Feng Miao\**

C. Pan, Y. J. Fu, J. W. Zeng, M. S. Long, E. F. Liu, C. Y. Wang, A. Y. Gao, M. Wang, Y. Wang, Prof. Z. L. Wang, Dr. S. J. Liang, Prof. F. Miao
National Laboratory of Solid State Microstructures, School of Physics, Collaborative Innovation Center of Advanced Microstructures, Nanjing University
Nanjing 210093, China
E-mail: miao@nju.edu.cn; sjliang@nju.edu.cn
J. X. Wang, Prof. R. Huang
Key Laboratory of Microelectronic Devices and Circuits (MOE), Institute of Microelectronics, Peking University
Beijing 100871, China

The current integrated circuit (IC) technology based on conventional MOS-FET (metal–oxide–semiconductor field-effect transistor) is approaching the limit of miniaturization with increasing demand on energy. Several analog circuit applications based on graphene FETs have been demonstrated with less components comparing to the conventional technology. However, low on/off current ratio caused by the semimetal nature of graphene has severely hindered its practical applications. Here we report a graphene/MoTe$_2$ van der Waals (vdW) vertical transistor with V-shaped ambipolar field effect transfer characteristics to overcome this challenge. Investigations on temperature dependence of transport properties reveal that gate tunable asymmetric barriers of the devices are account for the ambipolar behaviors. Furthermore, to demonstrate the analog circuit applications of such vdW vertical transistors, we successfully realized output polarity controllable (OPC) amplifier and frequency doubler.

These results enable vdW heterojunction based electronic devices to open up new possibilities for wide perspective in telecommunication field.

## 1. Introduction

With continuous scaling down of conventional MOSFET (metal–oxide–semiconductor field-effect transistor), integrated circuit technology is facing challenges on both miniaturization and power consumption. Several fundamental analog circuit functions based on graphene FETs[1-7] have been demonstrated with less components comparing to conventional FET technology, [8-12] suggesting potential applications on lowering the power consumption and diminishing the chip area. This is due to the unique ambipolar transfer characteristics of graphene based FETs. Nevertheless, the low on/off current ratio caused by a lack of a band gap in graphene severely hinders its practical applications. Among all scenarios to address this challenge, [13-27]van der Waals (vdW) heterostructures integrated by graphene and two-dimensional (2D) semiconductors [14-19, 21] have attracted enormous research interests due to potential high speed and flexible circuit applications. For these kinds of graphene-based vertical FETs (VFETs), high performance logic functions have already been realized, such as complementary inverters based on vertically stacked VFETs[19] and Ion-Gel-Gated VFETs.[21] However, graphene-based VFETs exhibiting ambipolar characteristics suitable for analog circuit applications have not been demonstrated.

In the design of graphene-based VFETs, searching for appropriate semiconductor materials is vital since their field effect transfer characteristics are mainly determined by the Schottky barrier (SB) formed at the graphene/semiconductor interface.[19] 2H-type molybdenum ditelluride ($MoTe_2$) is an air-stable layered semiconductor with suitable band gap near 1.1 eV for monolayer (~1.0 eV for bulk).[28, 29] Previous theoretical work suggested that the Fermi level pinning of $MoTe_2$ is at the middle of the band gap,[30] indicating both types of carriers (electrons and holes) can be effectively injected. Hence, 2H-type MoTe$_2$ offers an ideal material candidate for achieving desirable ambipolar transfer characteristics in graphene-based VFETs.

Here we report a graphene/MoTe$_2$/metal VFET (GM-VFET) showing suitable V-shaped ambipolar characteristics for possible analog applications. The mechanism responsible for V-shaped ambipolar charge transport is discussed, through the analysis of temperature-dependent carrier transport across asymmetric Schottky barriers tunbed by electrostatic field. Utilizing the unique feature of GM-VFET, we successfully demonstrated the output polarity controllable (OPC) amplifier and frequency doubler, which are two fundamental functions in analog circuits. Our work suggests that graphene/semiconductor/metal vertical heterostructure can be exploited as a promising platform to simplify the analog circuits design.

## 2. Results and Discussion

The schematic drawing and a representative false-color SEM image of a GM-VFET are shown in **Figure 1**a and Figure 1b respectively. After successful exfoliations of monolayer graphene and multi-layer MoTe$_2$ (around 20 nm) flakes on SiO$_2$ substrate, we used polyvinyl alcohol (PVA) transfer method to get stacked structure of graphene/MoTe$_2$ heterojunction.[31, 32] A standard electron beam lithography followed by metal electron beam evaporation processes was carried out to fabricate metal contacts (5nm Ag/40nm Au) (see details in expermental section). We also utilized Raman spectrum to characterize our devices with typical data presented in Figure 1c. The left and right figures are corresponding to the Raman spectra of MoTe$_2$ and graphene respectively, which are consistent with the previous studies.[33, 34] The channel length (thickness of the MoTe$_2$ flake can be measured by an atomic force microscopy (AFM) (see **Figure S1** in the Supporting Information).

We first examined the electrical performances of GM-VFETs, with room temperature (RT) small-bias output characteristics (drain-to-source current density ($J_{ds}$) vs drain-to-source voltage ($V_{ds}$)) of a representative GM-VFET shown in **Figure 2**a. Here different colors represent different applied back gate voltages ($V_{bg}$) ranging from -80V to 80V with a 20V interval. We did the measured current normalization based on the channel area, defined by the overlap area between the underlying graphene source and the top Ag/Au drain electrode. When increasing $V_{bg}$ from -80V, the current density was found to decrease first and reach a minimum value around -20V. After increasing $V_{bg}$ more towards the positive side, the current density starts to increase monotonously. This clearly indicates an ambipolar field effect behavior. Figure 2b shows the typical RT field effect transfer characteristics ($J_{ds}$-$V_{bg}$ curves) of a GM-VFET under different positive and negative bias conditions ($V_{ds}$=-0.05V, -0.1V, 0.05V, 0.1V). The maximum n-branch on/off current ratio of the GM-VFET is beyond $10^3$. The maximum p-branch current on/off ratio reaches ~20, which can be further enhanced through contact engineering such as using larger work function metals like Au (see Supporting Information Figure S2). To the best of our knowledge, this is the first demonstration of clear V-shaped ambipolar transfer characteristics in 2D VFET devices.

In contrast with the previously studied VFETs, we designed the top metal/semiconductor interface to be Schottky contact (rather than Ohmic contact), which plays an important role in the observed V-shaped ambipolar transfer characteristics in our devices. In our device structure, the Fermi levels of both graphene and MoTe$_2$ can be effectively tuned by the back gate. Thus, the transfer characteristics of the devices are largely determined by the gate tunable transport properties of both interfaces: the bottom graphene/MoTe$_2$ interface and the top MoTe$_2$/metal interface. To gain insight into the V-shaped ambipolar transfer characteristics as shown in Figure 2b, we drew four different band diagrams in Figure 2c to have a better understanding of the different carrier transport regimes, which are dependent on energy band bending at the two interfaces and tunablibity of Fermi level under the action of electrostatic fields. As we sweep the back gate voltage $V_{bg}$ from -80 V to 80 V for $V_{ds}$, we always see dips in the transfer curves at $V_{bg} = V_{Imin}$, which may be attributed to the intrinsic doping in the graphene and MoTe$_2$. For the regime I where both $V_{bg}$ and $V_{ds}$ are negative, the valence band and conduction band of MoTe$_2$ are bent upward, and the degree of band-bending is much stronger at the interface of

graphene/MoTe$_2$ than Au/MoTe$_2$, making the holes (as the majority carrier) injection from souce to drain easier than electrons injection through the barrier formed at the MoTe$_2$/metal interface. Thus the transport property of the device is dominated by holes. On the other hand, the positive $V_{bg}$ at the regime II makes the conduction and valence bands of MoTe$_2$ bend downward and the negative $V_{ds}$ leads to lower and thinner effective potential height for electrons (as majority carrier) through MoTe$_2$/metal interface than holes injection through the graphene/MoTe$_2$ interface. Consequently electron transport predominates the device performance. Similar to the above two regimes, the band diagrams for another two regimes (III and IV) with positive bias are also shown in Figure 2c. Note that the free-tuning of Fermi level in the graphene layer via electrostaic field enables more efficienct carriers (electrons or holes) injection through graphene/MoTe$_2$ interface than metal/MoTe$_2$ interface (see Figure S3 in the Supporting Information), due to the Fermi level pinning effect at the metal/MoTe$_2$ interface[30, 35].

To investigate the Schottky barriers in GM-VFET devices quantitatively, we further performed studies of temperature dependence of their electrical properties. The field effect transfer characteristics of a typical device at various temperatures ranging from 120 to 300K are shown in **Figures 3**a and 3b (with $V_{ds}$=0.2V and -0.2V respectively). Clearly, an increase in temperature leads to larger current. This indicates that the current injection through effective barrier height can be described by the thermionic emission model,[23, 28] in which the effective barrier height $q\phi_{eff}$ can be determined by using the temperature dependence of current $I \propto T^2 \exp(\dfrac{-q\phi_{eff}}{k_B T})$ (see Supporting Information S4 for details). Figure 3c shows the plot of $ln(I/T^2)$ versus $q/k_B T$ for various $V_{bg}$ with fixed $V_{ds}$ = 0.2V (see supporting information Figure S5b for the same analysis with $V_{ds}$ = -0.2V). Through fitting the date points at higher temperature regime (depicted by the dashed lines in Figure 3c), we can estimated the effective barrier height, with results plotted in Figure 3d. We found that the effective barrier height for $V_{ds} < 0$ is smaller than that for $V_{ds} > 0$, regardless of n-type doping or p-type doping. This means that the carrier injection (electron or hole) under $V_{ds} < 0$ is more efficient than that under $V_{ds} > 0$. It is worth to note that the extracted barrier height is the effective barrier height, considering that thermionic emission and tunneling are concomitant processes.[35, 38-40] At high temperatures, thermionic emission is dominant. But with further increasing $V_{ds}$, the effective barrier becomes thin, then the tunneling contribution becomes greater than thermioinc emission.

GM-VFET retaining such V-shaped ambipolar behaviors with high voltage gain is suitable for realizing fundamental analog circuit functions (see Figure S6 in the Supporting Information for comparison between graphene FET and GM-VFET). As a proof of concept, we demonstrated the realization of an OPC amplifier and a frequency doubler by connecting a GM-VFET in series with an off-chip load resistor. They are two fundamental analog circuit functions which are significant to simplify the circuits in common communications such as binary phase shift keying and binary frequency shift keying. To gain better gate tunability, we fabricated a GM-VFET with 30nm BN as gate dielectric (see Supporting Information Figure S7a). **Figure 4a** is the illustration of GM-VFET's field effect transfer curve for simplicity (see Supporting Information Figure S7b showing the transfer curve of the device). Figure 4b depicts the measurement circuit. The dynamic input signal was mixed with a sinusoidal signal $V_{ac}$ and a constant DC bias $V_{dc}$, and was applied to the gate electrode of GM-VFET ($V_{bg}=V_{ac}+V_{dc}$). The output signal was detected by an oscilloscope connected to the shared electrode of GM-VFET and the load resistor. We first demonstrated an OPC amplifier based on a GM-VFET, with results shown in figure 4c and 4d. In the measurement circuit, when GM-VFET exhibits as a p-type FET, for larger/smaller $V_{bg}$, more insulating/conductive device states and larger/smaller $V_{ds}$ will be achieved, yielding the same phase for the input and output signals, as shown in Figure 4c. On the other hand, when GM-VFET works as an n-type FET, $V_{ds}$ decreases with the increase of $V_{bg}$, which synchronously yields a phase difference of 180° of the output signal with respect to the input signal. The results are clearly shown in Figure 4d. In particular, both two operation modes we demonstrated above show nearly perfect phase match without unintended phase shift, suggesting high performance and great potential to develop more complicated analog circuit applications such as phase shift key.

The second fundamental analog circuit function we demonstrated is the frequency doubler, which requires the working bias of $V_{dc}$ setting at the minimum current point of the V-shaped ambipolar field effect transfer curve ($V_{dc} = V_{Imin}$). In **Figure 5a**, we schematically illustrate the working principle of such frequency doubler. The measurement circuit is the same as the OPC amplifier (depicted in Figure 4b). Points A to E in Figure 5a are the corresponding input (blue) and output (red) signal levels. When the input signal swings from point A to B, the GM-VFET acts as an n-type transistor, and the output signal $V_{ds}$ thus increases with decreasing input signal $V_{bg}$. While the input signal swings from point B to C, the GM-VFET acts as a p-type transistor, and the output signal $V_{ds}$ displays opposite variation tendency to the input signal $V_{bg}$. Similar to these two cases, the input and out signal swings from point C to E through D are also shown in the Figure 5a. It can be easily identified that each half-period swing of the input signal leads to a full-period swing in the output signal, thus realizing a frequency doubler. In Figure 5b, we show the excellent performance of GM-VFET-based frequency doubler without identified distortion. Based on our GM-VFET-based frequency doubler, binary frequency shift keying, a special case of frequency modulation with important applications in microwave radio and satellite transmission systems, could be realized with future research efforts.

## 3. Conclusion

In conclusion, the graphene/MoTe$_2$/metal vdW heterojunction-based field effect transistor has been fabricated with high quality. We studied the electrical transport properties of GM-VFET, and first demonstrated the V-shaped ambipolar field effect transfer characteristics in this structure via utilization of gate-tunable asymmetric barriers of the device. Through the analysis of temperature dependence of field effect transfer characteristics, the effective barrier height tuned by back gate voltage was estimated. Finally, the unique V-shaped ambipolar behaviors of GM-VFET enable us to successfully demonstrate OPC amplifier and frequency doubler based on a simple circuit of a single GM-VFET connected in series with an off-chip load resistor. Our findings offer a new pathway way for achieving future analog circuit applications based on vertical vdW heterostructure.

## 4. Experimental Section

*Device Fabrication*: We mechanically exfoliated graphene (Kish Graphite), MoTe$_2$ (HQ-graphene, Inc), and BN (HQ-graphene, Inc) respectively, and then transferred them onto SiO$_2$ (300nm)/Si substrate. The thickness of flakes was measured by using the Bruker Multimode 8 atomic force microscope (AFM). Our homemade micromanipulation system was utilized to stack the vertical heterostructures by PVA method [31, 32]. A standard electron-beam lithography process (FEI F50 with Raith pattern generation system) and electron-beam evaporation process were carried out to fabricate metal contact to graphene and MoTe$_2$. The device was then annealed at 280°C in Ar$_2$ ambient for 2 hours to remove resist residue.

*Electrical Characterization:* The electrical measurements of GM-VFET at RT were carried out in N$_2$ protecting ambient by an Agilent B1500A parameter analyzer. We conducted the variable-temperature measurements in an Oxford Instruments TeslatronTM CF cryostat, and the current signals were detected by Keithley 2635A System SourceMeter. For the measurement of OPC amplifier and frequency doubler, the AC source was generated by Stanford Research System SR830 DSP Lock-In Amplifier, and DC source was generated by Keithley 2400 SourceMeter. We detected the AC signal with Agilent DSO-X 2024A Digital Storage Oscilloscope.

**Supporting Information**
Supporting Information is available from the Wiley Online Library or from the author.

**Acknowledgements**

C. Pan and Y. J. Fu contributed equally to this work. This work was supported in part by the National Key Basic Research Program of China 2015CB921600 and 2013CBA01603, National Natural Science Foundation of China (61625402, 11374142 and 61574076), Natural Science Foundation of Jiangsu Province (BK20140017 and BK20150055), Fundamental Research Funds for the Central Universities, and Collaborative Innovation Center of Advanced Microstructures.

Received: ((will be filled in by the editorial staff))
Revised: ((will be filled in by the editorial staff))
Published online: ((will be filled in by the editorial staff))

**Figure 1 Graphene/MoTe₂ vertical stack VFET.** a) Schematic illustration of the GM-VFET. b) False-color SEM image of a typical fabricated device. Red dash line marks the edge of graphene. Scale bar is 5μm represented by red solid line. c) The left figure is the Raman spectra of MoTe$_2$, two peaks corresponding to the out-of-plan A$_{1g}$ mode at 172cm$^{-1}$ and the in-plan E$^1_{2g}$ mode at 232cm$^{-1}$. The right figure shows the Raman spectra of graphene, and the two most intense features correspond to the G peak at 1580cm$^{-1}$ and the 2D peak at 2700 cm$^{-1}$

**Figure 2 Room temperature (RT) transport characteristics of GM-VFET.** a) RT output characteristics of the GM-VFET under different $V_{bg}$ values. b) RT field effect transfer characteristics of the GM-VFET under different positive $V_{ds}$ (0.05V, 0.1V) values and negative V$_{ds}$ (-0.05V, -0.1V) values. c) Schematic band diagrams of GM-VFET corresponding to transfer characteristics in Figure 2b (I) $V_{ds}$ <0 and $V_{bg}$ < $V_{Imin}$, (II) $V_{ds}$ <0 and $V_{bg}$ > $V_{Imin}$, (III) $V_{ds}$ >0 and $V_{bg}$ < $V_{Imin}$ , and (IV) $V_{ds}$ >0 and $V_{bg}$ > $V_{Imin}$.

**Figure 3 Temperature-dependent charge transport of GM-VFET.** $I_{ds}-V_{bg}$ field effect transfer characteristics at different temperatures ranging from 120 to 300 K with $V_{ds}$=0.2V in a) and $V_{ds}$=-0.2V in b). c) Arrhenius plot at $V_{ds}$= 0.2V with $V_{bg}$ varying from –70 V to –50 V (p-branch) and 30 V to 70 V (n-branch). d) The variation of effective barrier height estimated from the slope of the fitted lines in c) and in Figure S3b of the Supporting Information as a function of $V_{bg}$.

**Figure 4 The demonstration of OPC amplifier.** a) The illustration of GM-VFET's field effect transfer curve. b) The schematic of measurement circuits, Both of R$_{load}$ and R$_0$ are 100 kΩ and the $C_0$ is 10nF. The frequency and amplitude of AC source is 1 KHz and 2V respectively. The $V_{dd}$ is 0.5V. c) When $V_{dc}$ is fixed at -5.9V, the phase of output signal well matches the input signal in the common-drain mode. d) With $V_{dc}$ is fixed at 1V, the output signal shows a phase difference of 180° with respect to input signal in the common-source mode.

**Figure 5 The demonstration of Frequency doubler.** a) The schematic diagram illustrates the basic working principle of GM-VFET based frequency doubler. b) The output signal shows frequency doubling compared to the input signal with $V_{dc}$ fixed at -2.8V.

**Figure 1**

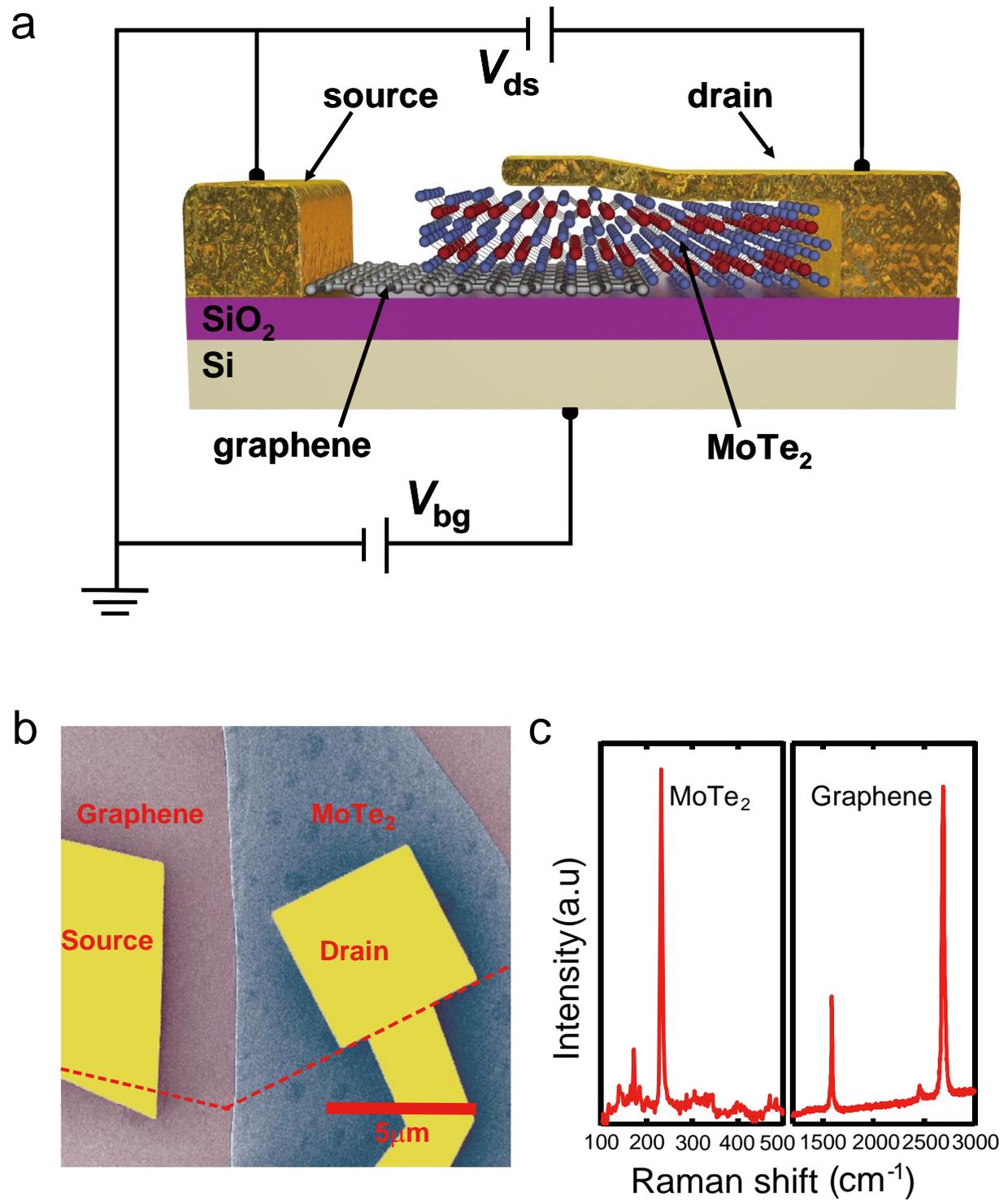

**Figure 2**

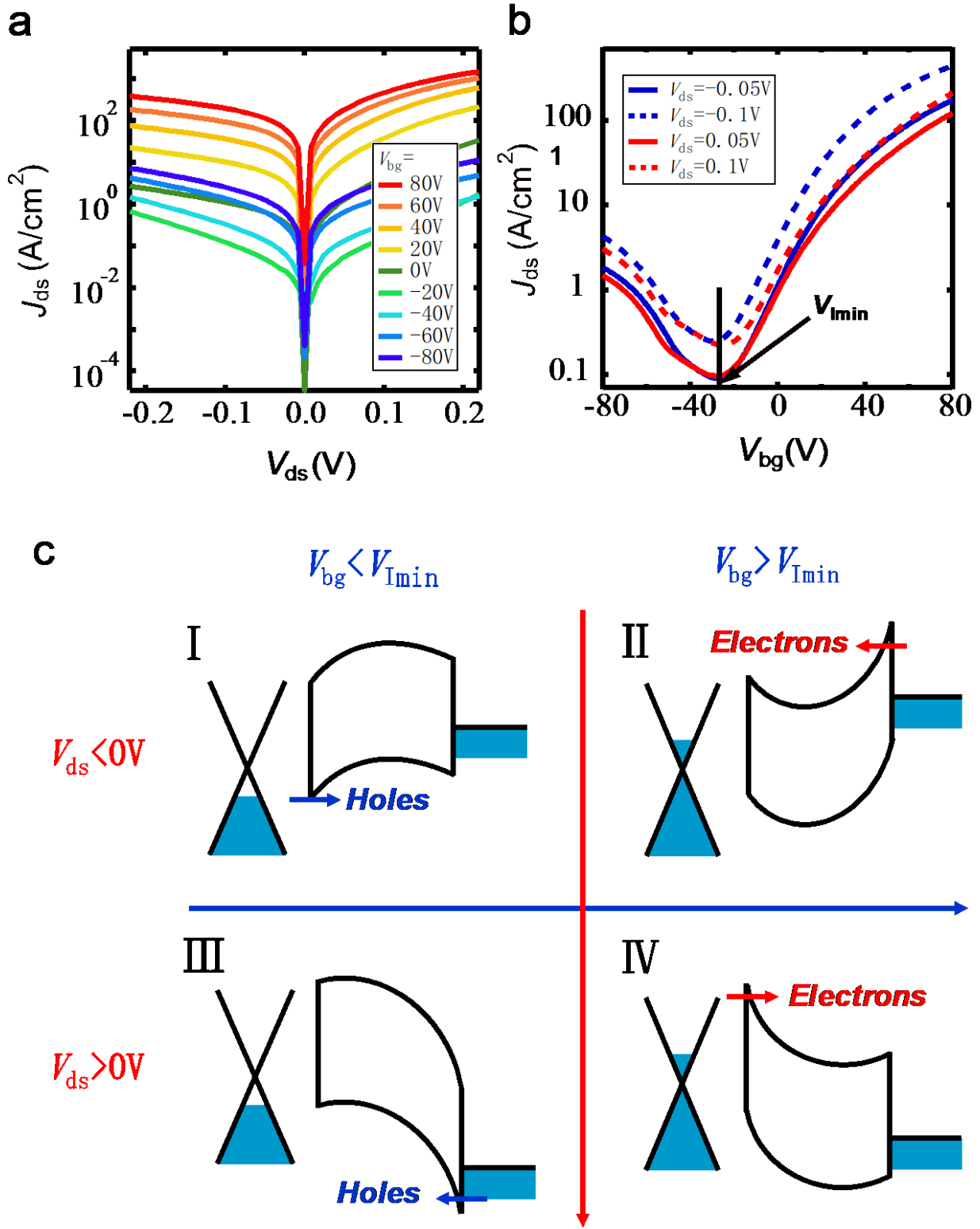

**Figure 3**

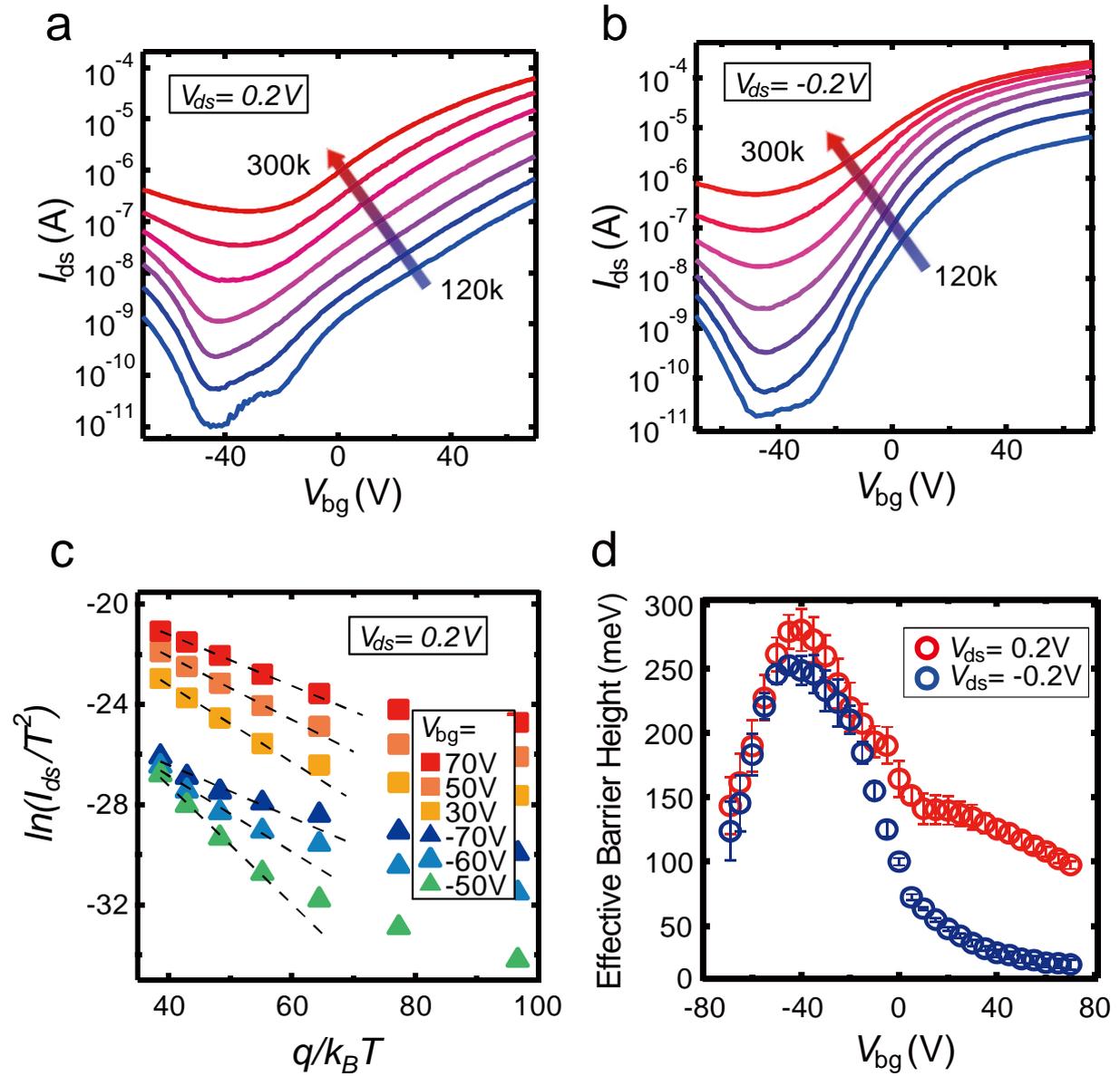

**Figure 4**

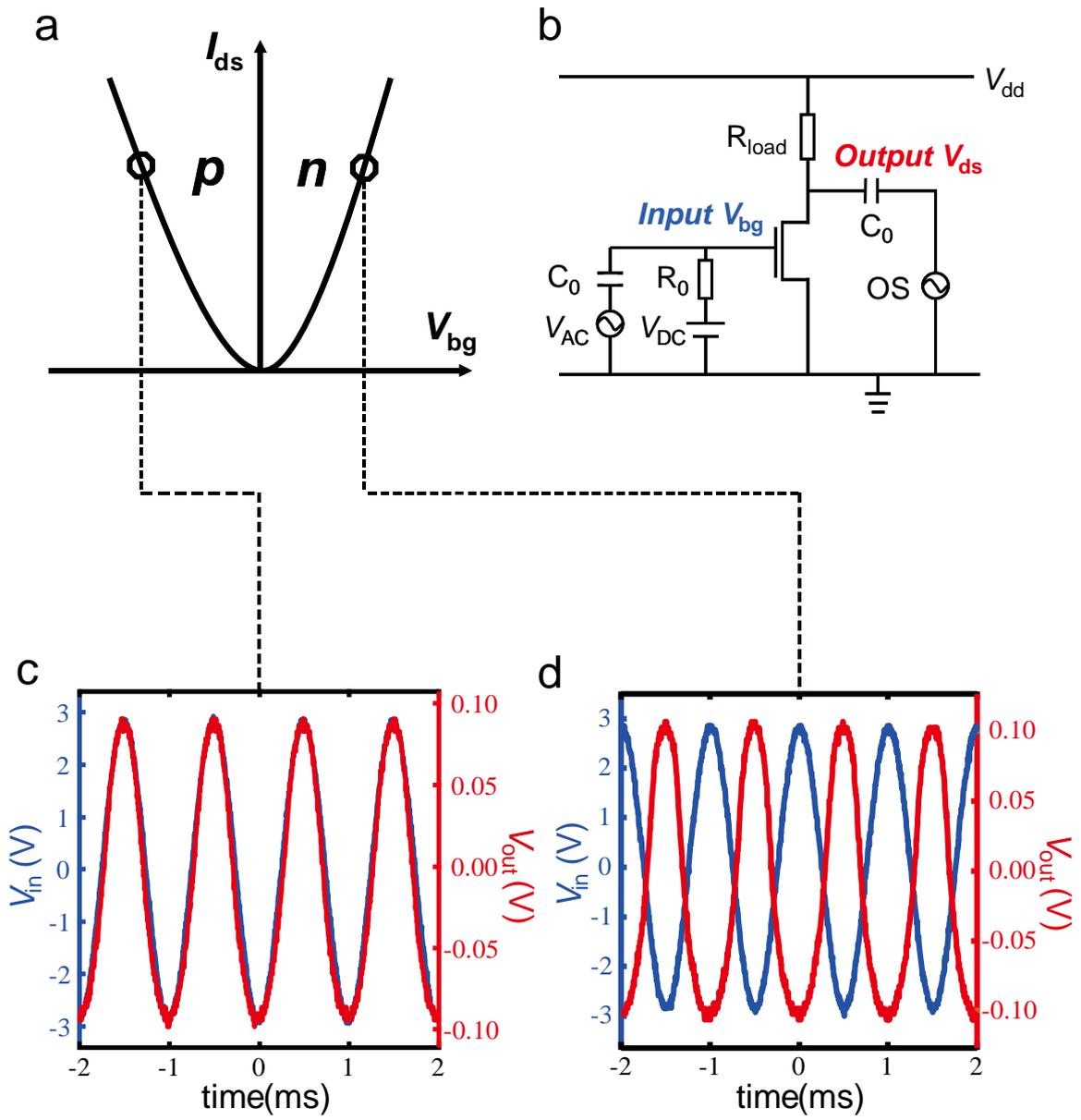

**Figure 5**

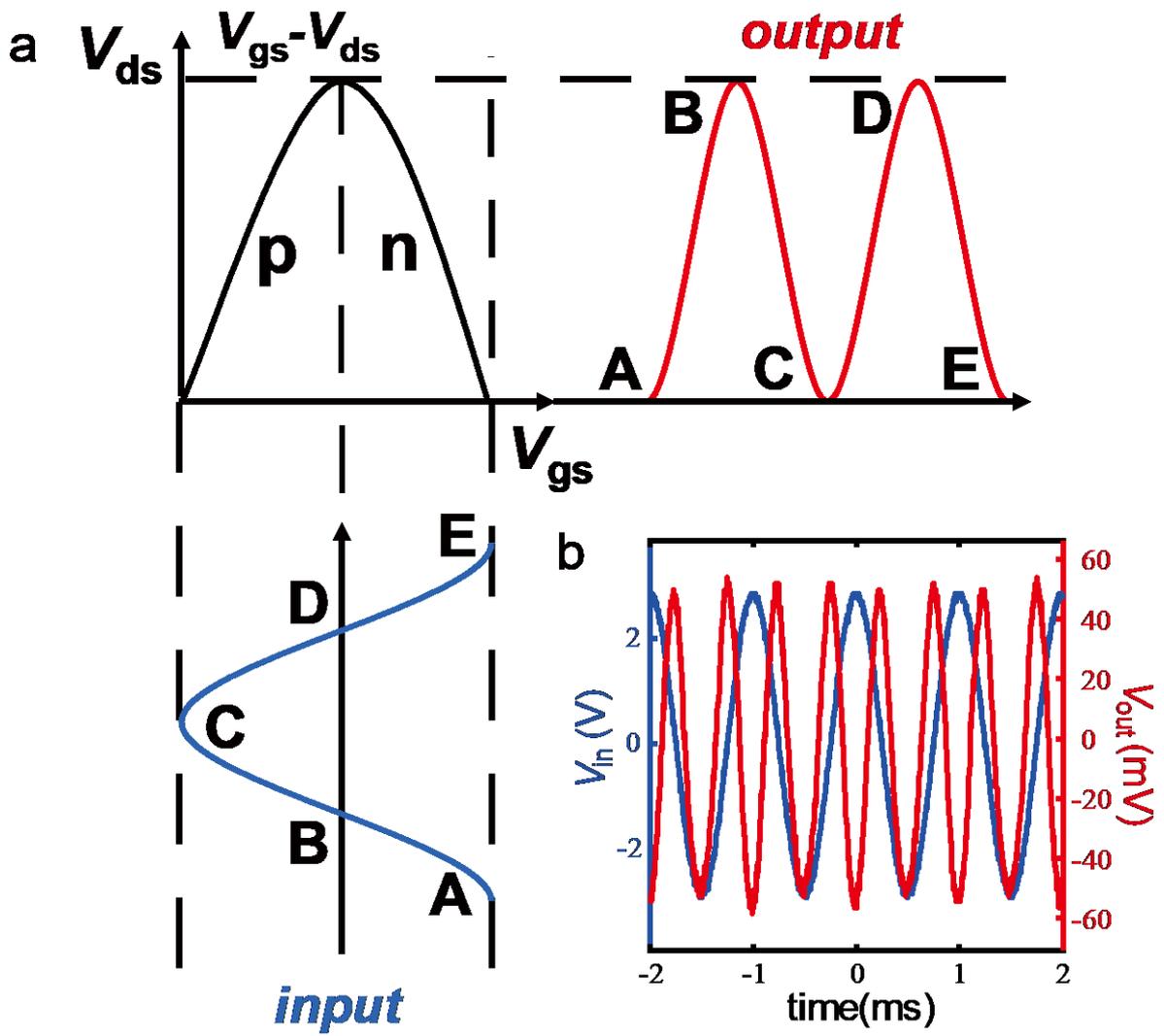

**V-shaped Ambipolar behaviors Graphene/MoTe2 Van der Waals Vertical Transistor** is first reported. Through detailed analysis of temperature dependent I-V characterstics, we attribute the V-shaped ambipolar behavior to gate tunable asymmetric barriers of the device. This approach offers us an opportunity to demonstrate high-performance output polarity controllable amplifier and frequency doubler, which enables graphene based electronic devices to open up a new possibility for wide perspective in telecommunication field.

**Keyword** vertical transistor, graphene/MoTe$_2$ heterostructure, ambipolar, output phase control, frequency doubler

C. Pan, Y. J. Fu, J. X. Wang, J. W. Zeng, G. Su, M. S. Long, E. F. Liu, C. Y. Wang, A. Y. Gao, M. Wang, Y. Wang, Shi-Jun Liang R. Huang, F. Miao

**Analog Circuit Applications based on V-shaped Ambipolar Behaviors of Graphene/MoTe$_2$ Vertical Transistors**

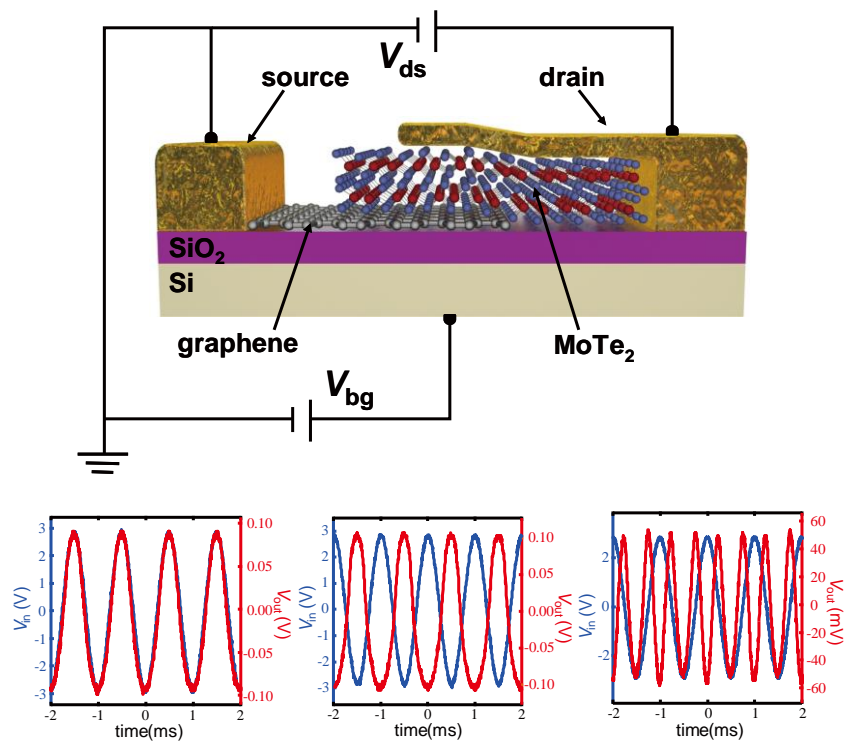

Supporting Information

**Analog Circuit Applications based on V-shaped Ambipolar Behaviors of Graphene/MoTe$_2$ Vertical Transistors**

*Chen Pan\*, Yajun Fu\*, Jiaxin Wang, Junwen Zeng, Guangxu Su, Mingsheng Long, Erfu Liu, Chenyu Wang, Anyuan Gao, Miao Wang, Yu Wang, Zhenlin Wang, Shi-Jun Liang, Ru Huang, Feng Miao*

**S1. The AFM image of graphene/MoTe$_2$ heterostructure**

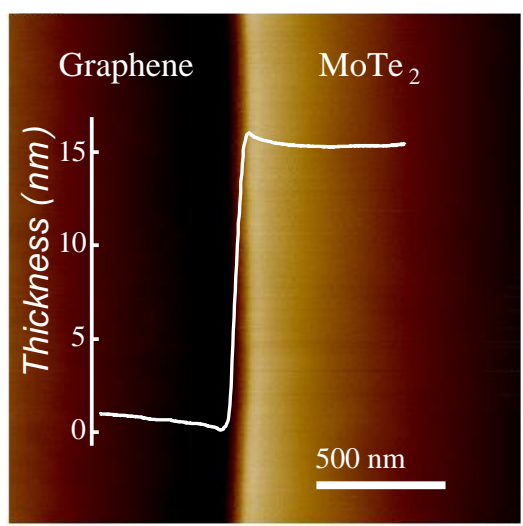

**Figure S1** The AFM image at the edge of the MoTe$_2$ flake. Inset: cross-sectional height profile of the MoTe$_2$ flake on the graphene/SiO$_2$/Si substrate. The thickness of the flake was estimated to be about 15.6 nm.

**S2. RT transfer characteristics of the GM-VFET with Au contact**

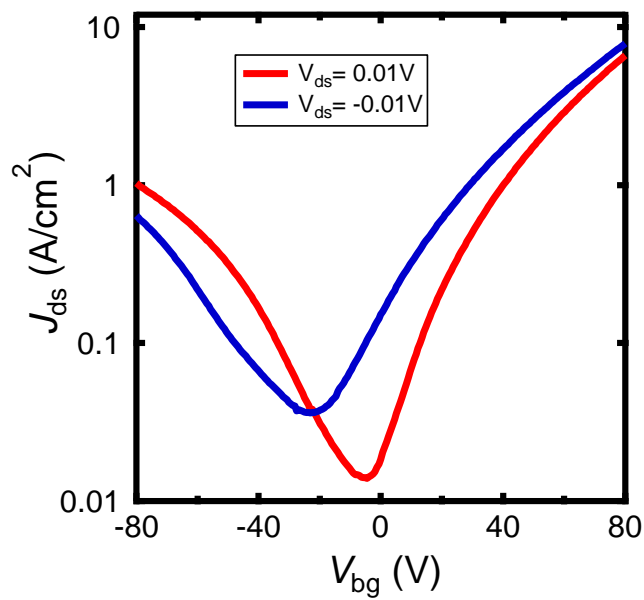

**Figure S2** RT transfer characteristics of the GM-VFET with Au as drain contact, which exhibit current on/off ratio of larger than 70 for p-branch and on/off ratio of lager than 450 for n-branch under $V_{ds}$=0.01V.

**S3. RT transfer characteristics of the GM-VFET with Few layer Graphene (~3nm) contact**

To reduce the Fermi level pinning effect at the interface of metal and $MoTe_2$, we also fabricated the GM-VFETs with few layer graphene as top metal contact. The schematic illustration and the optical microscopy image of GM-VFET are shown in the Figure S3-1. Compared to Ag/$MoTe_2$ interface, graphene(few layer)/$MoTe_2$ van der Waals heterostructure can effectively reduce the influence of Fermi level pinning on the carriers' transport.[1, 2] In the Figure S3-2, we show the RT transfer characteristics of the GM-VFET with few-layer Graphene (~3nm) as drain contact. Compared to Figure 3a and Figure 3b in the main text, the modulation of carrier current via gate voltage with few-layer graphene as drain contact becomes stronger than that with Ag as drain contact. This is due to lack of the Fermi pinning effect at the few-layer graphene/$MoTe_2$ interface, two Schottky barriers at the source and drain can be simultaneously tunbed by the applied gate voltage.

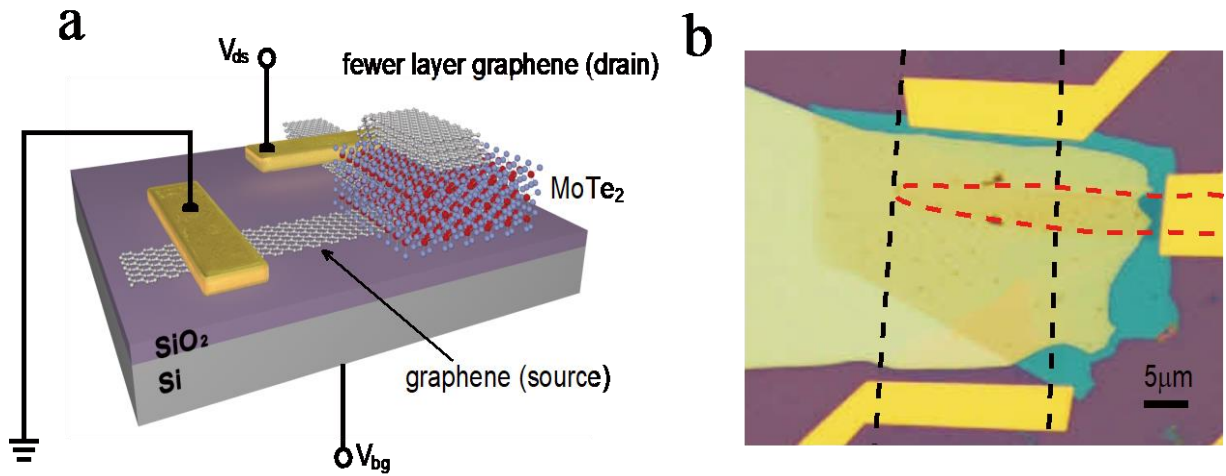

**Figure S3-1 a)** Schematic illustration of the GM-VFET with fewlayer graphene as drain contact. **b)** Optical microscopy image of the fabricated device. The underlying graphene is highlighted by the black dashed lines and the top few layer graphene is marked by red dashed lines. Scale bar (black solid line) is 5 μm. The top of the device is capped by h-BN.

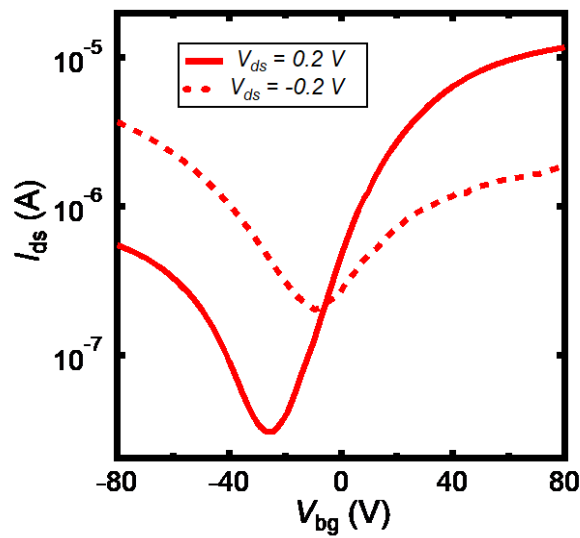

**Figure S3-2** RT transfer characteristics of the GM-VFET with few-layer graphene (~3nm) as drain contact for $V_{ds}$=0.2V (solid line) and $V_{ds}$= -0.2V (dash line).

### S4. Conventional thermionic emission model for effective barriers

GM-VFET device in the main text can be regarded as conventional metal–semiconductor–metal (M–S–M) structure.[2] Therefore thermionic emission model can be utilized to describe the carrier transport process:[3, 4] $I = AA^*T^2 \exp(\frac{-q\phi_{eff}}{k_BT})[\exp(\frac{qV}{nk_BT})-1]$, where $A$ is the area of the Schottky contact, $A^*$ is the effective Richardson constant, $q$ is the elementary charge, $k_B$ is the Boltzmann constant, $T$ is the Kelvin temperature, $V$ is the bias across source and drain, and n is the ideal factor. When SB is revisely-biased, $[\exp(\frac{qV}{nk_BT})-1] \approx -1$, the effective barrier height $\phi_{eff}$ can be determined by the temperature dependence of current $I \propto T^2 \exp(\frac{-q\phi_{eff}}{k_BT})$. Specifically, plotting $\ln(I/T^2)$ versus $q/k_BT$ gives a straight line, the slop for which is the effective Schottky barrier height.

**S5. Data for temperature dependent transfer characteristics**

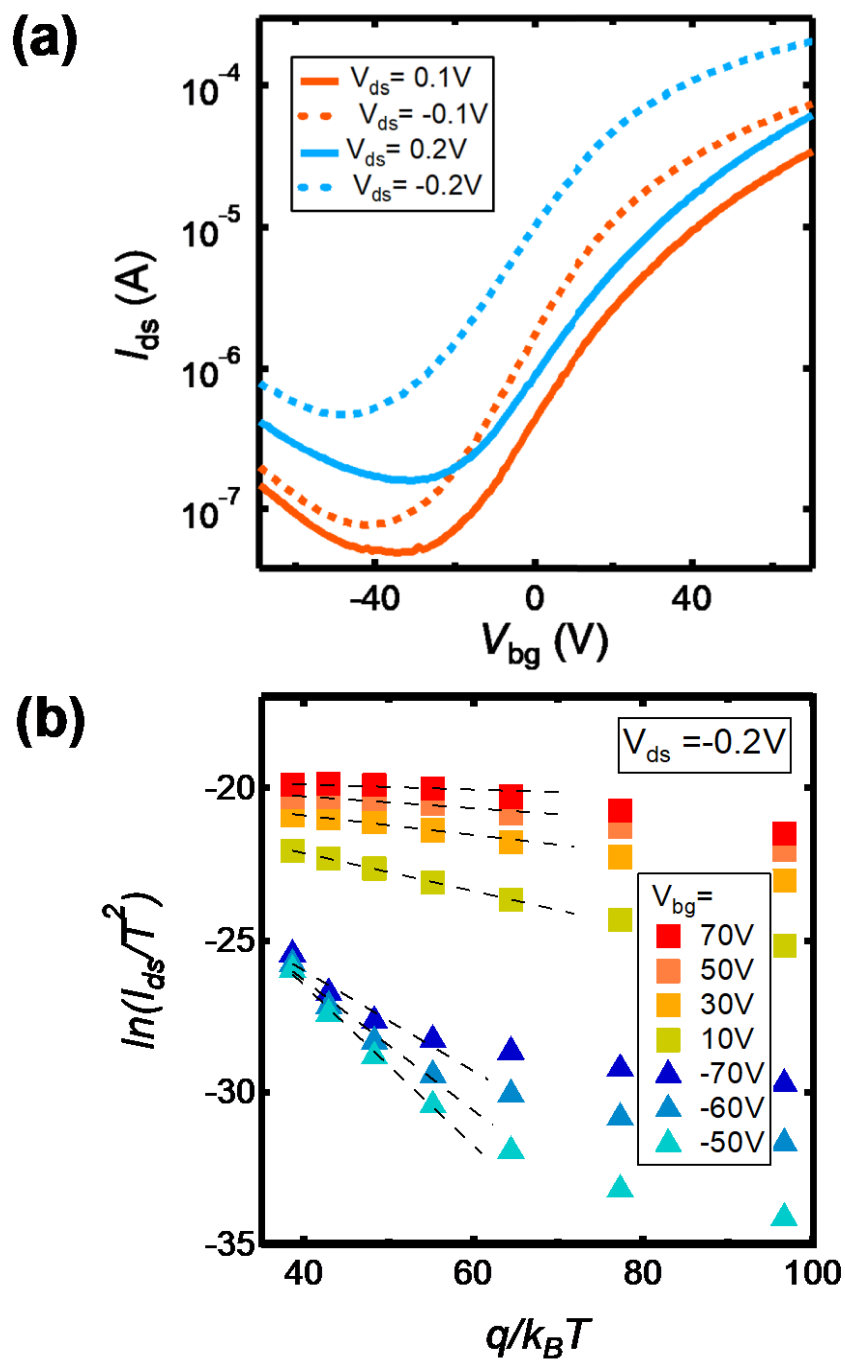

**Figure S5 (a)** Room temperature (RT) transport characteristics of the GM-VFET. **(b)** Arrhenius plot at $V_{ds}$=-0.2V with various $V_{bg}$ from -50 V to 70 V, and the best linear fit of the experiment data point over the high temperature regime can be done.

**S6. Comparison between graphene FET and GM-VFET**

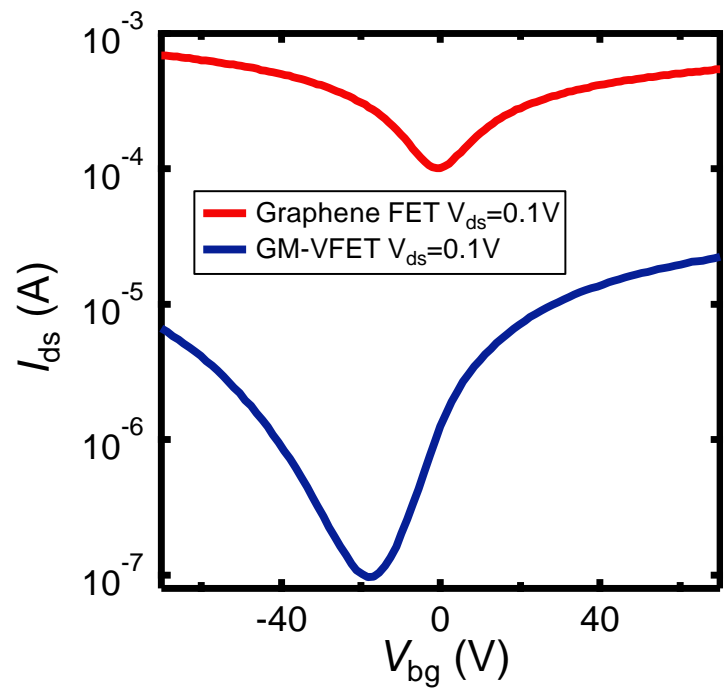

**Figure S6** The comparison between graphene FET and GM-VFET demonstrates that GM-VFET can amplify the on/off current ratio without sacrificing V-shaped ambipolar behavior, which is suitable for analog circuit applications.

**S7. The GM-VFET with BN dielectric**

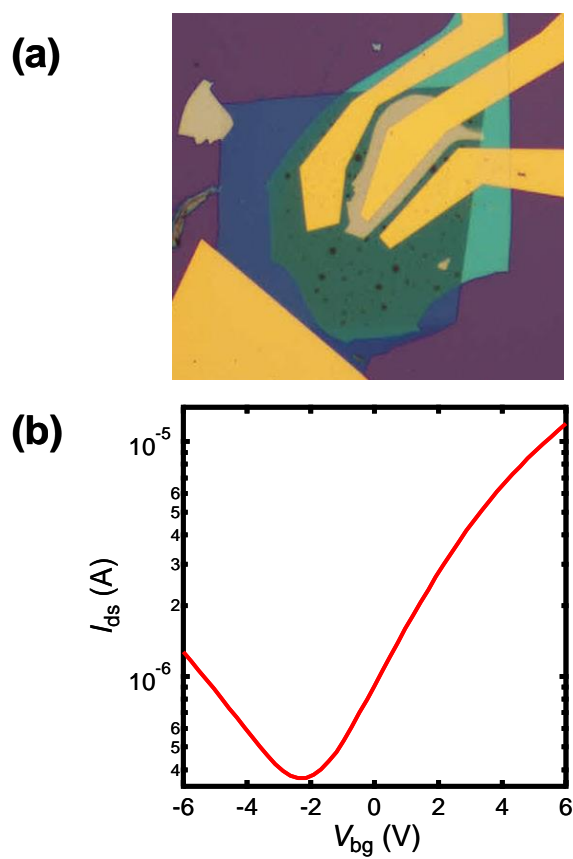

**Figure S7 (a)** Optical microscopy image of the fabricated device. We used ~20nm BN as gate dielectric and ~5nm graphite as gate electrode. **(b)** The transfer curve of the device for $V_{ds}$=0.2V.